\documentclass[12pt]{article}
\usepackage[a4paper, hmargin=2.5cm, vmargin=2.5cm]{geometry}
\usepackage{amsfonts,amsmath}
\usepackage[authoryear]{natbib} \bibpunct{(}{)}{,}{a}{,}{,}

\newcommand{\bI}{\mbox{\boldmath $I$}}
\newcommand{\bY}{\mbox{\boldmath $Y$}}
\newcommand{\by}{\mbox{\boldmath $y$}}
\newcommand{\bU}{\mbox{\boldmath $U$}}
\newcommand{\bzero}{\mbox{\boldmath $0$}}
\newcommand{\bmu}{\mbox{\boldmath $\mu$}}
\newcommand{\bDelta}{\mbox{\boldmath $\Delta$}}
\newcommand{\bLambda}{\mbox{\boldmath $\Lambda$}}
\newcommand{\bOmega}{\mbox{\boldmath $\Omega$}}
\newcommand{\bpsi}{\mbox{\boldmath $\psi$}}
\newcommand{\bSigma}{\mbox{\boldmath $\Sigma$}}
\newcommand{\btheta}{\mbox{\boldmath $\theta$}}
\newcommand{\bzeta}{\mbox{\boldmath $\zeta$}}
\newcommand{\vs}{\vspace{0.2cm}}

\begin{document}

\title{Comment on ``Hidden truncation hyperbolic distributions, finite mixtures
thereof, and their application for clustering''
by Murray, Browne, and \ McNicholas}

\author{Geoffrey J. McLachlan$^{1, \star}$, Sharon X. Lee$^2$}
\date{}

\maketitle

\begin{flushleft}
$^1$Department of Mathematics, University of Queensland, 
Brisbane, Queensland, Australia, 4072, Australia.\\
$^2$School of Mathematical Sciences, University of Adelaide, 
Adelaide, South Australia, 5005, Australia.\\
$^\star$ E-mail: g.mclachlan@uq.edu.au
\end{flushleft}

\begin{abstract}
We comment on the
paper of Murray, Browne, and McNicholas (2017),
who proposed mixtures of skew distributions,
which they termed hidden truncation hyperbolic (HTH).
They recently made a clarification (Murray, Browne, McNicholas, 2019) 
concerning their claim that the so-called CFUST distribution 
is a special case of the HTH distribution.
However, there are also some other matters 
in the original version of the paper
that are in need of clarification as discussed here.
\end{abstract}

\section{Introduction}
\label{sec:1}

In the paper Murray, Browne, and McNicholas (2017), herewith
referred to as MBM, consideration was given to mixtures of 
skew distributions belonging to the family
of distributions that
were termed hidden truncation hyperbolic (HTH).
One matter in MBM that has since been clarified by the authors
(Murray et al., 2019)
is that the CFUST distribution is not a special
case of the HTH distribution.
In the sequel, 
we wish to address some additional matters
in MBM that are also in need of clarification.

In particular,
we wish to point out that the comparison
in MBM with two other skew mixture models 
(called classical and SDB skew-$t$ mixtures in MBM) 
did not include mixtures of
canonical fundamental skew $t$ (CFUST) distributions.
As explained in Lee and McLachlan (2014),
an attractive feature of the CFUST distribution is that 
it includes the two distributions
corresponding to the classical and SDB skew $t$ distributions as special
cases, and it can be fitted with little additional effort;
see also Lee and McLachlan (2015, 2016).
We would also point out that Lee and McLachlan (2014, 2016)
have provided the EM equations for the fitting of a mixture
of CFUST distributions. 
Moreover,
Lee and McLachlan (2015, 2018) have given 
an R package for the fitting of a mixture of CFUST distributions.

It is not as if the CFUST distribution is a special case of the
HTH distribution as noted above.
It is to be demonstrated here that
the HTH distribution 
belongs to the family of skew distributions
that have the form of the canonical fundamental 
skew symmetric generalized hyperbolic (CFUSSGH) distribution.

We shall report here results showing that mixtures of CFUST distributions
outperform mixtures of HTH distributions for the two
real data sets considered in MBM. 

On another point on the comparison in MBM,
the near-zero value reported for the ARI of one of the
models in the comparisons of MBM
is due to algorithmic failure
rather than the claimed inability of the model to fit the data.


Before we consider further the aforementioned points, we define the skew 
distributions relevant to this paper.

\section{Skew distributions}
\label{subsec:skew}

To  establish some notation,
we let $\bY$ denote a $p$-dimensional random vector,
$\bI_p$ be the $p \times p$ identity matrix, 
and $\bzero$ be a vector/matrix of zeros of appropriate size. 
The notation $|\by|$ implies 
taking the absolute value of each element of $\by$.      
Also, we let
$\phi_r(\by;\,\mu,\bSigma)$ and
$\Phi_r(\by;\,\bmu,\bSigma)$ denote the $r$-dimensional
multivariate normal density and (cumulative) normal distribution
function, respectively,
with mean $\bmu$ and covariance $\bSigma.$

The skew distributions to be considered here
belong to the class of canonical fundamental skew symmetric (CFUSS)
distributions proposed by Arellano-Valle and Genton (2005).
The density of members of the class of CFUSS distributions can be 
expressed as 
\begin{eqnarray}
	f(\by;\,\btheta) &=&	2^{r} f_p(\by;\,\btheta) \, Q_r(\by;\,\btheta),  
\label{eq:100}
\end{eqnarray}  
where $f_p(\by;\,\btheta)$ is a symmetric density on ${R}^p$, 
$Q_r(\by;\,\btheta)$ is a skewing function 
that maps $\by$ into the unit interval, 
and $\btheta$ is the vector 
containing the parameters of $\bY$. 
Let $\bU$ be a $r\times 1$ random vector, 
where $\bY$ and $\bU$ follow a joint distribution 
such that $\bY$ has marginal density $f_p(\by; \btheta)$ 
and $Q_r(\by; \btheta) = P(\bU > \bzero \mid \bY = \by)$. 
If the latent random vector $\bU$ (the vector of skewing variables) 
has its canonical distribution
(that is, with mean $\bzero$ 
and scale matrix $\bI_r$), 
we obtain the canonical form of (\ref{eq:100}),
namely the CFUSS distribution.  
The class of CFUSS distributions encapsulates many existing distributions, 
including those to be considered here.
We now proceed to define those members.

\subsection{CFUSN distribution}
\label{subsec:cfusn}

The so-called canonical fundamental skew normal (CFUSN)
distribution is a location-scale variant of the 
canonical fundamental skew normal distribution in 
Arellano-Valle and Genton (2005).
If the $p  \times 1$ random vector $\bY$ has a CFUSN distribution,
its density is given by
\begin{equation}
f_{\rm CFUSN}(\by;\,\bmu, \bSigma,\bDelta)
=2^r \phi_p(\by;\,\bmu,\bOmega)
\,\Phi_r(c(\by);\,\bzero,\bLambda),
\label{eq:200}
\end{equation}
where
\begin{eqnarray*}
\bLambda&=&\bI_r\,-\,\bDelta^T\,\bOmega^{-1}\bDelta,\\
c(\by)&=&\bDelta^T\,\bOmega^{-1}(\by-\bmu),\\
\bOmega&=&\bSigma\,+\,\bDelta\bDelta^T, 
\end{eqnarray*}
and $\bDelta$ is a $p \times r$ matrix of skewness parameters with 
$pr$ free parameters.
This density is invariant under permutations of the columns of 
$\bDelta$, but this does not affect the number of free parameters.

The convolution-type characterization of this distribution
is given by
\begin{equation}
\bY=\bmu+ \bDelta |\bU_0| + \bU_1,
\label{eq:201}
\end{equation}
where
\begin{eqnarray}
\left[\begin{array}{c}\bU_0\\\bU_1\end{array}\right] \sim
N_{r+p} \left(\left[\begin{array}{c}\bzero\\\bzero\end{array}\right],\;
\left[\begin{array}{cc}\bI_r & \bzero\\\bzero & \bSigma\end{array}\right]\right).
\label{eq:203}
\end{eqnarray}

\subsection{CFUST distribution}
\label{subsec:cfust}

The canonical fundamental skew $t$ (CFUST) distribution can be characterized 
by
\begin{equation}
\bY=\bmu+ \bDelta |\bU_0| + \bU_1
\label{eq:204}
\end{equation} 
where, conditional on $W=w$,
\begin{eqnarray}
\left[\begin{array}{c}\bU_0\\\bU_1\end{array}\right] \sim
N_{r+p} \left(\left[\begin{array}{c}\bzero\\\bzero\end{array}\right],\;
w
\left[\begin{array}{cc}\bI_r & \bzero\\\bzero & \bSigma\end{array}\right]\right)
\label{eq:205}
\end{eqnarray}
and $W$ is 
distributed according to the inverse gamma distribution
$\mbox{IG}(\textstyle\frac{\nu}{2}, \textstyle\frac{\nu}{2})$.

If we take the matrix $\bDelta$ of skewness parameters
in the formulation (\ref{eq:204}) to be 
a $p$-dimensional vector (that is, $r=1$), then we obtain
the skew $t$-density as proposed by
Azzalini and Capitanio (2003).
Lee and McLachlan (2013a) referred to this distribution
as the restricted multivariate skew $t$ (rMST) distribution since 
restriction to a single skewing variable implies that
it is restricted to modelling skewness in only one direction 
in the feature space.
Sahu, Dey, and Branco (2003) considered the model where the 
matrix $\bDelta$ of skewness parameters was taken to be diagonal (so $r=p$).
Lee and McLachlan (2013a) termed this distribution the
unrestricted skew multivariate normal distribution to distinguish 
it from the rMST distribution, although they noted that it did
not embed the rMST distribution since
the feature-specific skewing effects are taken to be uncorrelated
(McLachlan and Lee, 2016).  
The rMST and uMST distributions are referred to as the classical and 
SDB skew $t$-distributions in MBM.
As they are special cases of the CFUST distribution
with the skewness matrix $\bDelta$
having $r=1$ and being diagonal, respectively,
we can denote them as CFUST($r$=1) and CFUST(diag), 
respectively.

Although the CFUST(diag) distribution can model skewness in multiple
directions, the assumption is that they are parallel to the axes 
of the feature space. 
Consequently, Lee and McLachlan (2014, 2015, 2016, 2018) developed
the methodology and algorithms for the fitting of mixtures of CFUST
distributions with arbitrary skewness matrices 
so that
they can handle skewness in multiple directions 
that are not necessarily
parallel to the axes of the feature space.

By letting the degrees of freedom go to infinity
in the formulation (\ref{eq:205}),
we obtain a similar formulation for the restricted multivariate
skew normal CFUSN($r$=1) and 
unrestricted multivariate skew normal CFUSN(diag) distributions.

\subsection{Scale mixture of CFUSN distribution}
\label{subsec:cfust}

A Scale Mixture of the CFUSN distribution (SMCFUSN)
can be defined by the stochastic representation 
\begin{eqnarray}
	\bY &=& \bmu + W^{\frac{1}{2}} \bY_0, 
\label{eq:206}
\end{eqnarray} 
where $\bY_0$ follows a central CFUSN distribution 
and $W$ is a positive (univariate) random variable 
independent of $\bY_0$. Thus, conditional on $W = w$, 
the density of $\bY$ has a CFUSN distribution 
with scale matrix ${w}\bSigma$. 
The marginal density of $\bY$ is given by
\begin{eqnarray}
&& f_{\mbox{\tiny{SMCFUSN}}} 
	(\by; \bmu, \bSigma, \bDelta; F_{\bzeta}) 
	\nonumber\\ && \hspace{0.5cm} = 
	2^{r} \int_0^\infty 
	\phi_p \left(\by; \bmu, {w}\bOmega\right) \, 
	\Phi_r\left(\frac{1}{\sqrt{w}}\bDelta^T\bOmega^{-1}(\by-\bmu); 
	\bzero, \bLambda \right) 
	dF_{\bzeta}(w), 
\label{eq:207}
\end{eqnarray}
where $F_{\bzeta}$ denotes the distribution function of $W$ 
indexed by the parameter $\bzeta$.  
We shall use the notation
$\bY \sim SMCFUSN_{p,r}(\bmu, \bSigma, \bDelta; F_{\bzeta})$ 
if the density of $\bY$ can be expressed in the form of (\ref{eq:207}).  

The CFUST distribution corresponds to taking $W$ to have 
the inverse gamma distribution 
$\mbox{IG}(\textstyle\frac{\nu}{2}, \textstyle\frac{\nu}{2})$.

\subsection{CFUSSGH distribution}
\label{subsec:cfusgh}

The so-called hidden truncation hyperbolic distribution in MBM can be
viewed as a member of the class of
the canonical fundamental skew symmetric generalized hyperbolic (CFUSSGH)
distributions (Lin, Lee, McLachlan, 2019).
To see this, we 
suppose now that the latent variable $W$ in (\ref{eq:206}) follows 
a generalized inverse Gaussian (GIG) distribution (Seshadri, 1997).
The GIG density can be expressed as 
\begin{eqnarray}
f_{\mbox{\tiny{GIG}}} (w; \psi, \chi, \lambda) 
	&=&	\frac{\left(\frac{\psi}{\chi}\right)^{\frac{\lambda}{2}} w^{\lambda -1}}
		{2 K_\lambda(\sqrt{\chi\psi})} 
		e^{-\frac{\psi w + \frac{\chi}{w}}{2}},
\label{eq:208}
\end{eqnarray}
where $W > 0$, the parameters $\psi$ and $\chi$ are positive, 
and $\lambda$ is a real  parameter. 
In the above, $K_\lambda(\cdot)$ denotes 
the modified Bessel function of the third kind of order $\lambda$. 
If we put $\chi=\nu$, $\lambda=-{\textstyle\frac{1}{2}}\nu$,
and let $\bpsi$ tend to zero in (\ref{eq:208}), 
then it tends to the inverse gamma distribution 
IG$({\textstyle\frac{1}{2}}\nu,{\textstyle\frac{1}{2}}\nu).$

Taking the latent variable $W$ in (\ref{eq:206}) to have a
GIG distribution, 
we obtain the CFUSSGH distribution.
It has the $p$-dimensional symmetric GH (generalized hyperbolic) density $h_p(\cdot)$
and the $r$-dimensional symmetric GH distribution function $H_r(\cdot)$,
corresponding to the symmetric density $f_p(\cdot)$ and the distribution
function $Q_r(\cdot)$, respectively, in (\ref{eq:100}).

The symmetric GH density is given by
\begin{eqnarray}
	h_p(\by; \bmu, \bSigma,\bpsi,\chi,\lambda)
&=&	\left(\frac{\chi+\eta(\by;\,\bmu,\bSigma)}
{\psi}\right)^{\frac{\lambda}{2}-\frac{p}{4}}
\frac{\left(\frac{\psi}{\chi}\right)^{\frac{\lambda}{2}} 
	K_{\lambda-\frac{p}{2}}(\sqrt{[\chi+\eta(\by;\,\bmu,\bSigma)]\psi})}
{(2\pi)^{\frac{p}{2}} |\bSigma|^{\frac{1}{2}} K_\lambda(\sqrt{\chi\psi})},\nonumber\\
&&\label{eq:209}
\end{eqnarray}  
where
$$\eta(\by;\,\bmu,\bSigma)=(\by-\bmu)^T\bSigma^{-1}(\by-\bmu).$$

It is well known that the GH distribution has an identifiability issue 
in that the parameter vectors $\btheta=(\bmu, k\bSigma, k\psi, \chi/k, \lambda)$
and $\btheta^*=(\bmu, \bSigma, \psi, \chi, \lambda)$ both yield 
the same symmetric GH distribution (\ref{eq:209}) for any $k>0$. 
It is therefore not surprising that the CFUSSGH distribution 
also suffers from such an issue. 
To handle this, restrictions are imposed on some of the parameters 
of the CFUSSGH distribution. 
An example is the HTH distribution considered in MBM
where the constraint $\psi=\chi=\omega$ is used, 
leading to the density
\begin{eqnarray}
&& \hspace*{-0.8cm} f_{\mbox{\tiny{HTH}}}
	(\by; \bmu, \bSigma, \bDelta, \omega, \lambda) 
	\nonumber\\ 
&& \hspace{-0.3cm} 
	= 2^r h_p\left(\by; \bmu, \bOmega, \omega, \omega, \lambda\right) 	
	H_r\left(\bDelta^T\bOmega^{-1}(\by-\bmu) 	
	\left(\frac{\omega}{\omega+\eta}\right)^{\frac{1}{4}}; 
	\bzero, \bLambda, \gamma, \gamma, 
	\lambda-\textstyle\frac{p}{2}\right), \nonumber\\
\label{eq:300}
\end{eqnarray} 
where $\gamma = \sqrt{\omega[\omega+\eta(\by;\,\bmu,\bSigma)]}$. 

Note that in their terminology, MBM are using `hidden truncation' 
to describe the latent skewing variable that 
follows a truncated distribution 
in the convolution-type characterization of the CFUSSGH distribution. 
Another alternative is to restrict the parameters of $W$ so that, for example, $E(W)=1$.  
A commonly used constraint on the GH distribution is to set $|\bSigma|=1$. 
This can be applied to the CFUSSGH distribution to achieve identifiability; 
see also the unrestricted skew normal generalized hyperbolic (SUNGH) distribution 
considered by Maleki, Wraith, and Arellano-Valle (2019).

The CFUST distribution is not a special case of the CFUSSGH distribution.
It can be obtained as a limiting case. 
One approach to obtain the limiting case is to put $\lambda=-\nu/2$ and to
replace $\bSigma$, $\bDelta$ and $\omega$ by
${\textstyle\frac{1}{k}}\bSigma, 
{\textstyle\frac{1}{k}}\bDelta$, 
and $k\nu$ in the density (\ref{eq:300}) to give
$$f_{\mbox{\tiny{HTH}}}
	(\by; \bmu,{\textstyle\frac{1}{k}}\bSigma,
{\textstyle\frac{1}{k}}\bDelta, k\nu, -\textstyle\frac{\nu}{2}),$$ 
and then to let $k$ tend to zero; see also Murray et al.\ (2019).

\section{Mixtures of CFUST distributions versus mixtures of HTH distributions}
\label{sec:cfustvhth}

To demonstrate the performance of mixtures of HTH distributions in clustering data,
they were compared in MBM with mixtures of the two special cases 
of CFUST distributions defined in the previous section, namely  the CFUST($r$=1) and 
CFUST(diag) distributions, the latter two being referred to as the
classical skew $t$ and the SDM skew $t$ distribution, respectively.
These three models were fitted to
two real data sets referred
to as the Seeds and HSCT (hematopoietic stem cell transplant) sets.
However,  a mixture of CFUST distributions was not fitted to these two data sets
in MBM, where it was 
stated that ``the SDB skew $t$ mixture model is regarded 
by some as the state of the art approach (see [36]).''
The reference [36] is the paper by Lee and McLachlan (2013a) 
in which the CFUST distribution was not considered. 
In subsequent papers (Lee and McLachlan, 2014, 2015, 2016),
it was explained and demonstrated how the SDB skew $t$ distribution,
along with the classical skew $t$, is embedded in the CFUST distribution. 
The results from MBM are reproduced here in Table 1, along with our results
for mixtures of CFUST distributions.
The notation HTHu and HTHm in Table 1 as used in MBM refer to the HTH
component distributions having $r=1$ and $r=p$ in forming
the skewness matrix $\bDelta$ in the formulation
of the HTH distribution
as a member of the class of SMCFUSN distributions (\ref{eq:207}).

On fitting  mixtures of CFUST distributions 
to the HSCT and Seeds data sets with $r=p$ skewing variables, we obtained 
higher values of the adjusted Rand Index (ARI), 
namely 0.991 and 0.916 
relative to the values of 0.984 (0.976) and 0.877 (0.877)
for mixtures of HTHm (HTHu) distributions
fitted to the HSCT and Seeds data sets, respectively.

\begin{table}[htbp!]
\caption{ARI value of CFUST mixture model versus the 
ARI values for the mixture models in MBM}
\begin{tabular}{|cc| |c| c c c c|}
\hline
& & CFUST& HTHu & HTHm & Classical skew $t$ & SDB skew $t$\\ \hline \hline
HSCT& &0.991 & 0.976   & 0.984  & 0.782&0.890  \\ \hline
Seeds & &0.916 & 0.877  & 0.877  & 0.836 & 0.009 \\ \hline
\end{tabular}
\end{table}

\vspace{.05cm}

In MBM, the very small value of 0.009 for the 
ARI for the SDB skew $t$ mixture model in Table 1 is not interpreted
as reflecting the failure of the algorithm to fit the model to the data.
Rather it is taken at face value with the statement that
``the SDB skew-$t$ mixture performs better than the 
classical skew-$t$ mixture approach for the Seeds data.''

As stated in MBM, ``the expected value of the ARI
under random classification is zero.''
Thus the reported value of 0.009 for the SDB skew $t$-mixture  model
is implying that this model
does no better than random classification!
But this is unrealistic, particularly as an ARI value of 0.836 was obtained
for the classical model; such a difference between the ARI's for these two models
is highly unlikely as both embed the $t$-mixture model with just a few additional
parameters to allow for any  skewness in the data.

To investigate this further,
we followed the procedure adopted in MBM 
for the fitting of this model (McNicholas, 2015).
We first scaled the data so that each variable
had mean zero and unit standard deviation
before applying the EMMIXuskew program (Lee and McLachlan, 2013b) 
using the default options for the starting values for the skewness parameters.
We found that the EM algorithm stopped after three iterations as
essentially  all the observations were being put into the one cluster.
It was a consequence of the initial
estimates of the skewness parameters not being scale invariant which for the
Seeds data set after scaling caused problems.
However, the EMMIXuskew algorithm  also has optional
starting strategies in addition to the default ones, 
such as those provided by $k$-means applied to the unscaled data
and by the normal mixture model.
When we used these latter options, we obtained a fit for the 
CFUST(diag) mixture model (that is, the SDB skew $t$ model)
with an ARI of 0.84.
However, 
given the problems that our algorithm EMMIXuskew encountered 
with the data as scaled in MBM,
we have modified those default steps for the provision of starting values
that were not scale invariant.
But we recommend using 
our latest algorithm EMMIXcskew (Lee and McLachlan, 2015, 2018),
which also has the provision to fit mixtures of CFUST distributions. 
\vs
\vs

\noindent
{\bf References}
\vs

\noindent
Arellano-Valle, R.B., and Genton, M.G. (2005). On fundamental skew distributions.
{\it Journal of Multivariate Analysis\/} {\bf 96}, 93--16.
\vs

\noindent
Azzalini, A. and Capitanio, A. (2003). Distributions generated by perturbation of
  symmetry with emphasis on a multivariate skew $t$ distribution. Journal of
{\it Royal Statistical Society  B} {\bf 65}, 367--389.
\vs

\noindent
Lee, S.X. and McLachlan, G.J. (2013a).
On mixtures of skew normal and skew $t$-distributions. 
{\it Advances in Data Analysis and Classification\/} {\bf 10}, 241--266.
\vs

\noindent
Lee, S.X. and McLachlan, G.J. (2013b). 
EMMIXuskew: an {R} package
for fitting mixtures of multivariate skew $t$-distributions via the EM
algorithm. 
{\it Journal of Statistical Software\/} {\bf 55}, {\it No.} 12.
\vs


\noindent
Lee, S.X. and McLachlan, G.J. (2014).
Maximum likelihood estimation
for finite mixtures of canonical fundamental skew $t$-distributions: the
unification of the unrestricted and restricted skew $t$-mixture models.
arXiv:\ 1401.8182v1 [stat.ME], 31 Jan 2014.
\vs

\noindent
Lee, S.X. and McLachlan, G.J. (2015). 
EMMIXcskew: an {R} package for the
fitting of a mixture of canonical fundamental skew $t$-distributions.
arXiv:\ 1509.02069.v1 [stat.CO], 7 Sept 2015.
\vs

\noindent
Lee, S.X. and McLachlan, G.J. (2016). 
Finite mixtures of canonical fundamental skew $t$-distributions: 
the unification of the restricted and unrestricted skew $t$-mixture models.
{\it Statistics and Computing\/} {\bf 26}, 573-589.
\vs

\noindent
Lee, S.X. and McLachlan, G.J. (2018).  EMMIXcskew: an {R} Package 
for the fitting of a mixture of canonical fundamental skew $t$-distributions. 
{\it Journal of Statistical Software\/} {\bf 83}, {\it No.} 3.
\vs

\noindent
Lee, S.X., Lin, T.-I., and McLachlan, G.J. (2018). 
Mixtures of factor analyzers with fundamental skew symmetric distributions.  
arXiv:\ 1802.02467v1 [stat.ME], 7 Feb 2018. 
\vs

\noindent
Maleki, M., Wraith, D., and  Arellano-Valle, R.B. (2019).
Robust finite mixture modeling of multivariate unrestricted
skew-normal generalized hyperbolic distributions.
{\it Statistics and Computing\/} {\bf 29}, 415--428.
\vs

\noindent
McLachlan, G.J. and Lee, S.X. (2014). 
Comment on ``Comparing two formulations
of skew distributions with special reference 
to model-based clustering" by
A.\ Azzalini, R.\ Browne, M.\ Genton, and P.\ McNicholas. 
{\it Statistics \& Probability Letters\/} {\bf 116}, 1-5.
\vs

\noindent
McNicholas, P.D. (2018). Private communication.
\vs

\noindent
Murray, P.M., Browne, R.B., and McNicholas, P.D. (2017).
Hidden truncation hyperbolic
distributions, finite mixtures thereof, and their application for
clustering.
{\it Journal of Multivariate Analysis\/} {\bf 161}, 141--156.
\vs

\noindent
Murray, P.M., Browne, R.B., McNicholas, P.D. (2019).
Note of clarification on ``Hidden truncation hyperbolic
distributions, finite mixtures thereof, and their application for
clustering,'' by Murray, Browne, and McNicholas,
J.\ Multivariate Anal.\ 161 (2017) 141--156.
{\it Journal of Multivariate Analysis\/} {\bf 171}, 475--476.
\vs

\noindent
Sahu, S.K., Dey, D.K., and Branco, M.D. (2003). 
A new class of multivariate skew
distributions with applications to Bayesian regression models. 
{\it Canadian Journal of Statistics\/} {\bf 31}, 129--150.
\vs

\noindent
Seshadri, V. (1997). Halphen's laws. 
In {\it Encyclopedia of Statistical Sciences\/}, 
S.\ Kotz, C. B.\ Read, and D.L.\ Banks (Eds.). 
New York: Wiley, pp.\ 302--306.

\end{document}